# Estimating the 3D Time Variable Water Vapor Contents of the Troposphere from a Single GNSS Receiver

Jean-Pierre Barriot[1], Jonathan Serafini[1], and Lydie Sichoix[1]
[1]Geodesy Observatory of Tahiti, University of French Polynesia, 98702 Faa'a, Tahiti, jean-pierre.barriot@upf.pf

## ABSTRACT

We describe here a new algorithm to model the water contents of the atmosphere from GNSS slant wet delays relative to a single receiver. We first make the assumption that the water vapor contents are mainly governed by a scale height (exponential law), and secondly that the departures from this decaying exponential can be mapped as a set of low degree 3D Zernike functions (w.r.t. space) and Tchebyshev polynomials (w.r.t. time.) We give an example of inversion with data acquired over a one day time span at the Geodesy Observatory of Tahiti.

### I. INTRODUCTION

The principal limiting error source in Global Navigation Satellite Systems (GNSS) and other modern space geodesy techniques (like Very Large Base Interferometry (VLBI) or Deep Space Probes tracking), is the mismodeling of the delay experienced by radio waves in propagating through the electrically neutral atmosphere, usually referred to as the troposphere delay [1]. The troposphere slant delay is generally split into a dry (hydrostatic) and a wet components. Each one can be described as a product of the delay at the zenith and a mapping function, which models the elevation dependence of the propagation delay, with sometimes the addition of azimuthal gradients [2]. In this paper we focus on the modeling of the slant wet delay (SWD), as it remains as the most problematic propagation effect affecting both range and Doppler measurements by radio waves [3]. On the other side, the vertical structure of water vapor in the atmosphere [4] is one of the initial information of numerical weather forecast models [5, 6, 7] and climate modeling [8]. Studies undergone at the European Space Agency (ESA) in the frame of the Bepi-Colombo mission to planet Mercury [9,10] indicate that measurements taken at a single site give good calibration of wet delays for long time periods (> 1,000 s), but not accurate enough for shorter times, and that water vapor radiometers (WVR) must be used to efficiently calibrate the wet delay over such time spans. We show in this paper that some improvement can be made to the current modeling of the wet delay from a single receiver by introducing a true 3D tomographic "mapping function" instead of the mapping functions of today.

### II. WATER VAPOR TOMOGRAPHY

*A. Water vapor tomography from a swarm of GNSS receivers*

The tomography of the water vapor contents of the atmosphere from slant wet delays has been a popular subject of research in the last decade, with tens of published papers (e.g. [6,11,12].) At the core of these algorithms is the inversion of the so-called Radon transform, i.e. the integral of the wet atmospheric refractivity along the line-of-sight rays between the receiver and the GNSS satellite. The high spatial and temporal variations of the troposphere demands a very large set of path wet delays along all possible line-of-sight directions coming from high density ground networks of GNSS receivers, with at least a 30 s sampling period. Typically, the zone of interest of the troposphere (100 to 1,000 km$^2$ on the ground up to 15 km altitude) is divided into blocks called voxels (a few hundred of meters in all directions), with a time resolution at around 30 minutes. Each voxel must be crossed at nearly the same moment by several rays to obtain a meaningful imaging. The intrinsic reason for that is that the Radon transform does not care about the position of the fluctuation of refractivity along the ray, but the fluctuation of the refractivity must be the same at a given point crossed (at nearly the same time) by two different rays.

*B. Water vapor tomography from a single GNSS receiver*

If we want to use a single receiver, we have of course to overcome these limitations. This type of inverse problem has been intensively studied in other fields, like medical imaging, where it is known as the fan-beam tomography problem [13]. Physically, the ill-posedness comes from the fact, already stated, that the radial part of the spatial structure of the refractivity is lost in the integration along the line-of-sight. Besides, by definition, we do not have any crossing rays. The only way to regain this lost information is to take into account, during the inversion process, *a priori* information about the expected spatial and temporal structure of the refractivity. At some stage, we will have to correlate, explicitly or implicitly, the reconstructed refractivity from ray to ray. Our algorithm is also based on the inversion of the Radon transform, which can be written as

$$\delta d(\vec{u}, t) = \int_0^L N(s\vec{u}, t) ds \quad (1)$$

where the delay $\delta d$ and $s$ are in the same units (see Fig. 1), $N$ is the wet refractivity, $\vec{u}$ is the line-of-sight unit vector, $t$ is the time of the measurement (we neglect the propagation time), $s$ is an affine parameter along the ray, from the GNSS receiver (0) to the top of the troposphere ($L$). From this point, we have to plug as much as physical information as we can into the system. In order to enforce the uniqueness of the reconstructed $N$ field we then write

$$\delta d = N_G \int_0^L \exp(-h/H + \xi) ds \quad (2)$$

where $h$ is the altitude, $H$ is an *a priori* water vapor scale height. Le positiveness of $N$ is expressed by the exponential, $N_G$ being the refractivity at the GNSS receiver location. The term $\xi$ reflects the departure of the refractivity from the decaying exponential and is the quantity we are looking for.

If we set $\xi = 0$ we end up with the well-known relation

$$\delta d_0 = N_G L \int_0^1 \exp\left(-\frac{\lambda L \sin(e)}{H}\right) d\lambda \sim \frac{N_G H}{\sin(e)} \quad (3)$$

If we assume that $\xi$ must be close to zero, we can use a first-order approximation and write

$$\delta d = N_G \int_0^L \exp(-h/H)(1+\xi) ds \quad (4)$$

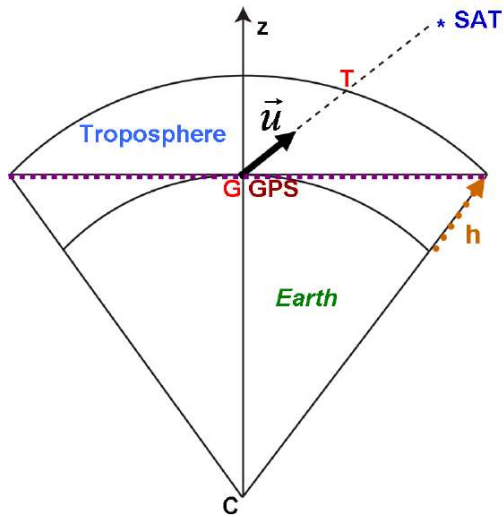

Figure 1.    *Geometry of a single station GNSS tomography. The distance L in Eq. 1 corresponds to the segment GT. The unit vector $\vec{u}$ is the line-of-sight vector (elevation and azimuth.) We assume in this paper that the tropopause is a spherical shell.*

We have to stress that this assumption means that we are looking for perturbations that are proportional to the decaying exponential, an assumption that is valid in the mean for the troposphere, but can be violated from time to time in the case of strong inversion layers.

The next step is to develop $\xi$ along a suitable basis of 3D function (w.r.t. space.) We choose the 3D Zernike basis, which is an orthonormal basis on the unit ball (and over the half-sphere by symmetry) of the form

$$R_n^l(\lambda) Y_l^m(\vec{u}) \quad (5)$$

where the functions $R_n^l(\lambda)$ are the set of 3D Zernike polynomials with $0 \leq \lambda \leq 1$, the functions $Y_l^m(\vec{u})$ are the set of usual (complex) harmonic functions over the unit sphere. The Zernike functions use three indexes, the driving index being $n$ (up to a given degree $n_{max}$), with $0 \leq l \leq n$ and $-l \leq m \leq l$. The precise definition of the 3D Zernike polynomials and 3D Zernike functions is highly technical, and we refer to [14,15] for further details. To take into account time variations, we join to this set an orthonormal basis relative to time, in the form of Tchebyshev polynomials (with a maximum degree $k_{max}$.) The advantage of Tchebyshev polynomials with respect to "naive" sine/cosine functions is that no periodicity pattern w.r.t time is assumed. We end up with

$$\delta d = N_G L \int_0^1 \exp\left(-\frac{\lambda L \sin(e)}{H}\right)(1 + \xi(\lambda L \vec{u})) d\lambda \quad (6)$$

$$\xi = \sum_{n,l,m} \sum_k E_{n,l,m}^k R_n^l(\lambda) Y_l^m(\vec{u}) T_k(\tau) \quad (7)$$

where $\tau$ is a reduced time that varies from -1 to 1 (the abscissa span of Tchebyshev polynomials) during the acquisition process. It is clear that the actual number of functions taken from this whole set must be kept to a bare minimum.

For Eq. (6) we use a rectilinear approximation for the rays. We refer to [16] for the validity of this assumption. Essentially, the delay variations caused by the bending of the rays are of second-order w.r.t. variations caused by the fluctuations of refractivity integrated over the rays (if the rays are not too close to the horizon.) Finally we obtain

$$\delta d - \delta d_0 = N_G L \int_0^1 \exp\left(-\frac{\lambda L \sin(e)}{H}\right) \xi(\lambda L \vec{u}) d\lambda \quad (8)$$

The technical difficulty here is to integrate the 3D Zernike polynomials along the rays. This can be done analytically with the ansatz

$$\int \exp(-a\lambda) \lambda^n d\lambda = \frac{1}{a^{n+1}} \Gamma(n+1, a\lambda) \quad (9)$$

where $\Gamma$ stands for the incomplete Gamma function of the third kind [17] $a = -\frac{\lambda L \sin(e)}{H}$, and $\lambda^n$ is any of the monomes of the 3D Zernike polynomials.

Casted this way, we now have a typical linear inverse to solve for, that can be heuristically written as $Gp = d$, with the vector $d$ gathering all the SWD observables during a given period of time, $p$ is the set of $E_{n,l,m}^k$ coefficients, and $G$ corresponds to the sensitivity matrix linking $p$ to $d$. To obtain a tractable problem, the usual way is to reduce its dimensionality by considering the set of normal equations $G^T G p = G^T d$, but this does not reduce the ill-posedness, which is of physical origin. To obtain a physically acceptable solution (i.e. a reasonably "smooth" refractivity field), we have to confine the spatial spectrum of the solution to the low portion of the frequency space. This can be done in essentially three ways: (*i*) by using low degrees $n_{max}, k_{max}$ in the series (7), (*ii*) by adding additional information, generally in the form of *a priori* covariance matrices, (*iii*) by confining the solution to a subset in proper spaces (i.e. by using a truncated Singular Value Decomposition (SVD) of the normal matrix). There is a lot of literature devoted to inverse problems in the frame of geophysics, and we refer to [18].

The best way to obtain a physically acceptable solution is of course to add physical information coming from other sources. This is already done, as we hard-wired an exponential decay in the wet refractivity field we are looking for (Eq. 2), after a lot of unsuccessful trials with more general forms. *A priori* covariance matrices can in principle be built from atmosphere turbulence theory [19] or empirically [20, 21], but this is extremely tricky, with very different length scales [22, 23], and so additional parameters. Time and space refractivity fluctuations can also be linked through a frozen flow atmospheric turbulence assumption [24], with an *a priori* knowledge of the wind velocity. To avoid introducing too many under-constrained *a priori* parameters, we choose to stay simple and opted for straightforward truncated singular value regularization. An optimal value for the truncation was derived by balancing the norm of the residuals w.r.t. the norm of the solution (Morozov's principle, see [25] for literature and a typical application.)

### III. APPLICATION TO REAL DATA

*A. Data set used*

We used a set of 2102 SWD points computed by post-processing residuals from a batch of GPS measurements taken at the Tahiti IGS THTI station [26] on October 15, 2007 UTC (ellipsoidal WGS84 coordinates: altitude 98 m, latitude 17.57 S, longitude 149.61 W.) This particular date is at the beginning of the rainy season in the Tahiti Island (mild tropical maritime climate.) The GIPSY-OASIS II software was used for this data processing in PPP mode at 5 minutes sampling [27, 28]. We stress that these reconstructed SWD integrates the line-of-sight residuals left over after data processing and were not edited. Then an *a priori* value of the refractivity at the surface was computed from humidity, pressure and temperature measurements at ground level [29, 30], and an *a priori* estimate of the water vapor scale height was derived from the SWD data set by applying the rule of thumb $\delta d_0 \sim \frac{N_G H}{\sin(e)}$ (Eq. 3.) Only data points with an elevation greater than 15° were taken into account. We found a value of $H = 2.4$ km. The thermal tropopause itself was fixed at a 16.5 km altitude with a mean temperature of – 75 °C (typical values given by Meteo-France.) After several trials, we selected $n_{max} = 4$ and $k_{max} = 32$, leading to 1155 unknowns to solve for. But this is only an apparent number of unknowns, as we only selected the first 200 leading singular values by balancing the norm of the residuals w.r.t. the norm of the solution, as previously stated, confining the power spectrum to low frequencies. The degree of data fit, defined as the complement to one of the norm of the vector of residuals divided by the norm of the data vector was 95 %. Outliers (less than 0.2 %) where rejected by applying a 3 $\sigma$ rule.

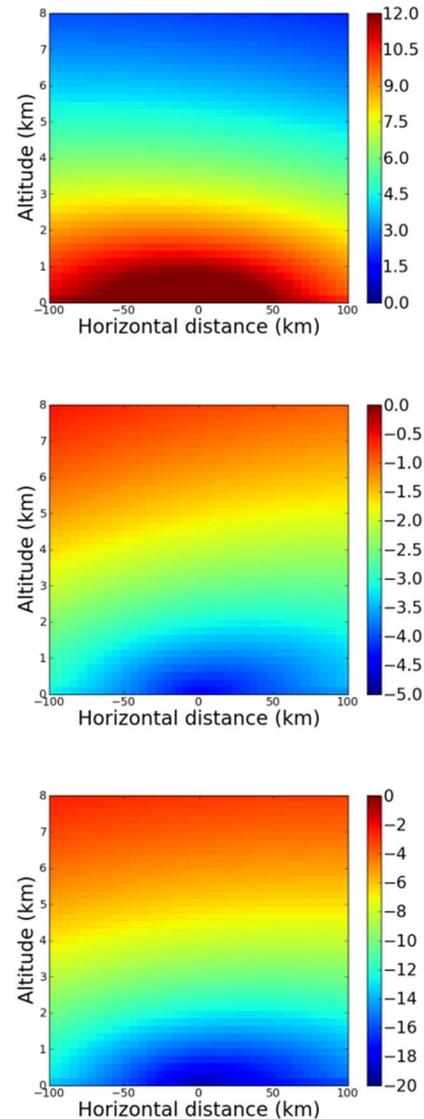

Figure 2. Wet refractivity variations in N-units (mm/km) at 00:00 h, 12:00 h and 24:00 h (top to bottom) on Oct. 15, 2007 (10:00h am to 10:00h am local time) w.r.t. an exponential decay obtained by applying the algorithm described in this paper. These times corresspond to the time of the radiosoundings (see Figure 4. The quantity mapped is $N_G \exp(-h/H) \xi$ with $H = 2.4$ km (see Eq. 4). The IGS station THTI is located at the (0,0) coordinates in the figure. The horizontal distance lies in the plane of the local horizon along the North-South azimuth. The local horizon crosses the tropopause at a distance of 466 km (limited at 100 km on the figures.) N-units scales are different on each sub-figure to enhance visible details. Ground mean relative humidity: 85 %, ground mean temperature: 26 °C, tropopause mean temperature: - 75°C at 16.5 km altitude. We estimate that the formal relative error is at the 10 % level.

*B. Results*

Figures 2 (wet refractivity variation maps), 3 (PW estimates) and 4 (comparisons with radiosoundings) illustrate our results. Figures 2 show that the water vapor repartition with altitude follows the layering evidenced by Lowry et al. [31]. In Figure 3, we computed two PW estimates: (yellow line) by converting our ZWD estimate (Zenith Wet Delay obtained by integrating the wet refractivity along the vertical) to PW by the multiplicative constant Pi [32], an empirical constant function of the atmospheric virtual temperature (here Pi had a mean value of 0.162); (red line) by integrating along the vertical the our water vapor density estimate derived from the wet refractivity by assuming a linear variation of the temperature from the ground to the tropopause. The black line represents the PW estimate derived from the GIPSY-OASIS ZWD by following the IGS model for tropospheric products [33]. One can see that the yellow and black lines are pretty close. This shows that our algorithm matches the performance of one of the best available GPS processing package (comparisons with PW estimates computed by using the BERNESE and GAMIT-GLOBK software are similar, see [26].) We stress that the PW estimates corresponding to the yellow and black lines are derived from their respective underlying ZWD estimates by multiplication of the same Pi constant. The red line represents the PW estimate that we computed by integrating along the vertical the water vapor density (with the same linear decay of the temperature as for the Pi constant). The red line presents a nearly constant negative bias of around 2.5 kg/m$^2$ with respect to the yellow line. We think that this bias is due to the fact that the Pi constant is computed by assuming an underlying hydrostatic law for the density of the water vapor, an assumption that is relaxed by definition when we integrate the water vapor density along the vertical. These PW estimates are compatible with the PW estimates derived from the radiosondes (humidity and temperature measurements) that are launched twice daily from the neighboring Faa'a airport (4 km), but with a strong bias that cannot be explained by the difference in altitude (92 m), and are inconsistent with the PW estimates derived from the humidity and temperature measurements of the surface weather station collocated with the GPS receiver at the Geodesy Observatory of Tahiti. Such biases between radiosounding and GPS estimates for large PW values in tropical areas have already been noted in the literature [34, 35] and can probably be explained both by the volume averaging resulting from the randomly distributed GPS satellites used and radiosonde sensor biases. Besides, we have to keep in mind that the ascent phase of the balloon before burst at an altitude of 22 km takes around 90 minutes with a lateral drift up to 100 km. In order to make comparisons possible with the radiosounding profiles, we have multiplied the wet refractivity computed from radiosondes measurements (see [36] for the procedure) by a 82 % factor. Figure 4 show two radiosoundings separated by 24 h and the corresponding tomography results. The radiosounding done at 00:00h UTC (10:00h am local time) shows a very strong inversion layer at 7 km altitude. In this case the inversion algorithm overestimates the wet refractivity in the lower troposphere to compensate for the inversion layer. For the radiosounding done at 24:00 UTC (10 am local time the following day), with no inversion layer present, the inversion algorithm is able to recover the wet refractivity structure, as we are now complying with the physical assumption of low departure from the *a priori* decaying exponential .

IV. CONCLUSION

We showed that, even in the case of a single GNSS receiver, we can go beyond the modeling of the wet delay by the usual mapping functions. There is no "magic" in the algorithm: only two basic assumptions: (i) we are looking for refractivity perturbations that are small with respect to an exponential decay of the refractivity with altitude, (ii) the refractivity perturbations are correlated from ray to ray with respect to their line-of-sight and temporal separations. We plan in the near future to do extensive tests of the algorithm through comparisons with tomography of the wet refractivity by GNSS with crossing-rays [11] and Raman Lidar [37, 38]. It is also clear that the algorithm will beneficiate of large constellations of GNSS satellites (GPS, GLONASS, Beidou…) that will reduce the volume averaging. We plan also to take into account the bending of the rays at low elevations by following the work of Marini [39].

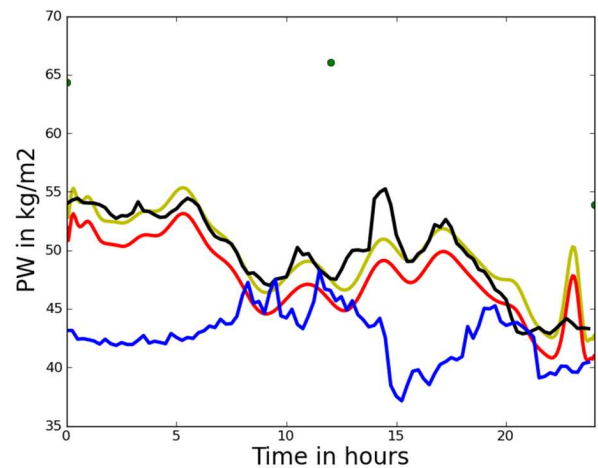

Figure 3.   Time variation of the precipitable water (PW) in kg/ m$^2$ at the THTI GPS station on Oct. 15, 2007 UTC. Red line: estimate computed by integrating along the vertical the water vapor density obtained by applying the algorithm described in this paper (see Figure 4.) Yellow line: Estimate computed by integrating along the vertical the wet refractivity (Zenithal Wet Delay) and converting it to PW by the multiplicative constant Pi (see text.) Black line: PW estimate derived from GIPSY-OASIS processing. Blue line: PW estimate from ground measurements of the weather station colocated with the GPS receiver, computed by assuming a water vapor scale of 2.4 km. Black dots: PW measurements from Meteo-France radiosoundings at the nearby Faa'a airport (64.32 kg/m$^2$ at 0:00h UTC, 66.03 kg/m$^2$ at 12:00h UTC, 53.86 kg/m$^2$ at 00:00h UTC the following day.)

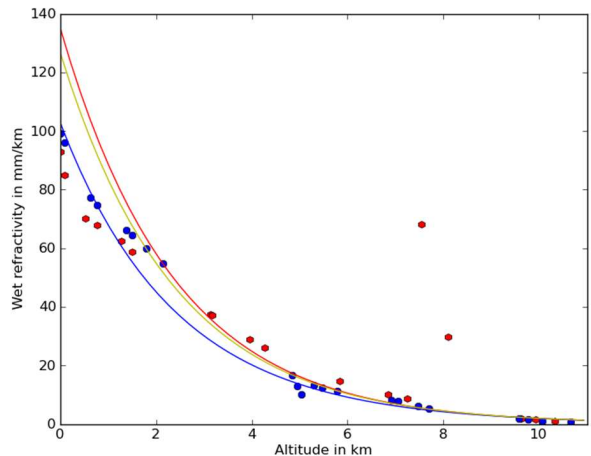

Figure 4. Comparison between two radiosoundings and the tomography results. Red dots: radiosounding done at 00:00h UTC, showing a very strong inversion layer at 7 km altitude. Red line: corresponding tomography, that overestimates the wet refractivity in the lower troposphere to compensate for the inversion layer. Blue dots: radiosounding done at 24:00h UTC, with no inversion layer present. In this case, the algorithm is able to recover the wet refractivity structure, as we are complying with the physical assumption of low departure from the *a priori* decaying exponential (yellow line). The radiosounding at 12:00h UTC is similar to the radiosounding at 00:00h UTC and not shown for clarity reasons.

ACKNOWLEDGMENTS

This research has been funded by the "Centre National d'Etudes Spatiales", under continuous grants covering the years 2011-2013 ("Décision d'Aide à la Recherche".) The PhD grant of J. Serafini was funded by the French ministry of Research (MENRT.) SWD residuals where computed by A. Fadil during his post-doc internship in the Geodesy Observatory of Tahiti. We also thank Victoire Laurent from Meteo-France for help with the radiosoundings.